\documentclass[12pt]{article}
\usepackage{amsmath,amsfonts,amssymb,graphicx,psfrag, epsfig,rotate}
\newcommand {\I}      {\'{\i}}

\newcommand {\Kp}     {{\mathbb K}}
\newcommand {\lp}     {{\rm 1\hskip-0.09cm l}}

\newcommand {\sump}   {\sideset{}{^{\prime\prime}}\sum}

\newcommand {\bra}   {\langle}
\newcommand {\ket}   {\rangle_{\hskip -0.05cm\rm }}
\newcommand {\ketg}  {\rangle_{\hskip -0.05cm\rm g}}

\newcommand{\pf}[1]{\psfrag{#1}{\small {#1}}}
\def\1barra{1\! \hskip -1.1pt {\rm l}}

\begin{document}


\title{ A Comment on the $\beta$-expansion of $s=\frac{1}{2}$ and $s=1$
Ising Models}
\author{Winder A. Moura-Melo$^1$, Onofre Rojas$^2$,
 E.V. Corr\^ea Silva$^3$,  \\
S.M. de Souza$^1$  and M.T. Thomaz$^3$\vspace{0.25cm} \\
\small\it
$^1$Departamento de Ci\^encias Exatas, Universidade Federal de
Lavras\\\small\it Caixa Postal 37, CEP: 37200-000, Lavras-MG,  Brazil
\vspace{0.25cm} \\
\small\it $^2$Departamento de F\I sica, Universidade
Federal de S\~ao Carlos \\\small\it CEP: 13565-905, S\~ao Carlos - SP,
Brazil. \vspace{0.25cm} \\\small\it $^3$Instituto de F\I sica,
Universidade Federal Fluminense \\\small\it Av. Gal. Milton Tavares de
Souza s/n.$\!\!^\circ$, CEP: 24210-340, Niter\'oi-RJ, Brazil}


\maketitle

\begin{abstract}
The purpose of the present work is to apply the   method recently developed
in reference \cite{chain_m} to the spin-1 Ising chain, showing how to
obtain analytical $\beta$-expansions of thermodynamical functions
through this formalism. In this method, we do
not solve any transfer matrix-like equations. A comparison
between the $\beta$-expansions of
the specific heat and the magnetic susceptibility for the
$s=\frac{1}{2}$ and $s=1$ one-dimensional
Ising models is presented. We show that those
expansions have poorer convergence when the auxiliary function
of the model has singularities.
\end{abstract}

\newpage


\sloppy

\section{Introduction}

\ \indent\hspace{-3mm} One dimensional (1D) lattice models have been widely
 studied for
decades, motivated, for example, by the possibility of
obtaining physical insights about more realistic theories and of
their utilization as good {\em laboratories} for testing the
applicability of new methods.\\

Several instances of chain models have been intensively
investigated, mainly by virtue of their relevance in connection with
Condensed Matter phenomena; among them, we can mention the one-dimensional
(1D) Ising, the spin-$1/2$ XXZ Heisenberg\cite{takahashi_L} and the
Hubbard\cite{Hubbard,klumperetc} models.   Actually, the chain models
mentioned above belong to the class of exactly solvable models, due
to their integrability property\cite{takahashi_L}. Thus, the algebraic
 Bethe
{\em ansatz} method has been widely applied in order  to determine their
thermodynamical functions by means of non-linear integral equations.

Recently\cite{chain_m},  Rojas {\it et al.} showed how to
obtain a closed analytical expansion for the grand potential in the thermodynamic
limit for any translationally invariant chain model from the high
temperature  expansion of the cumulant method\cite{domb}.  Another point
that we can stress  is that the method developed in ref.\cite{chain_m}
can also  be extended to models including nearest
interacting neighbors (e.g. frustrated quantum   Heisenberg models with
spin-S\cite{bonner,ivanov}).
For such quantum systems, the calculation of higher order terms in the
$\beta$-expansion of thermodynamical functions allows one to study their
properties at lower temperatures\cite{domb}.
Namely, very good
results have been obtained applying the method to the
(integrable) spin-1/2 XXZ chain model\cite{chain_m}. Actually, we should
stress that the applicability of such method extends further to
 either integrable or non-integrable models (the XXZ with
spin-1-case being an example of the latter case), provided that
the hamiltonian presents translational invariance, nearest-neighbor
interaction, and periodic boundary conditions.\\


In this paper we intend to obtain an analytical $\beta$-expansion of the
 Helmholtz free energy for the
spin-1 Ising model through the application of the results of reference
\cite{chain_m}. Using the fact that this model is exactly solvable,
the correctness of such expansion can be checked.
The derivation of the Helmholtz free energy for any unidimensional
chain model in the thermodynamical limit is
obtained from the auxiliary function $\varphi(\lambda)$, the latter
representing the actual mathematical improvement in this approach, in
relation to the literature (cf.\cite{domb}).
Moreover, the present paper is a first
step towards the application of such approach to 1D Ising
models with arbitrary spin\cite{workinprog1}, as well as to more interesting
models, such as the spin-1 XXZ Heisenberg chain
\cite{workinprog2}.\\


The importance and usefulness of the $\beta$-expansion of
thermodynamic functions relies on the perspective of obtaining information
 for
lower temperatures, as terms of higher order
in the expansion are taken into account. Hence, knowing how fast or how
 slow
such expansion converges is crucial to its applicability. We want to
clarify how the characteristics (namely, the singularities) of the
auxiliary function $\varphi(\lambda)$ impact on the convergence of the
$\beta$-expansion of any thermodynamic function. In order to
do so, we have studied the $\beta$-expansion of the thermodynamic functions
for the spin-1 and spin-1/2 Ising models.
At this point, we should stress once more that our $\beta$-expansion
 results
are analytic, non-perturbative and valid, in principle, for
arbitrary values of the parameters.

In section 2 we present a summary of the results derived in reference
 \cite{chain_m}.
In section 3 we obtain an analytical expression for the auxiliary function
$\varphi (\lambda)$ for the spin-1 Ising model for finite values of
 $\beta$;
in section 4 we discuss the rapidity of the convergence of the
 $\beta$-expansion of two
thermodynamical functions, namely, the specific heat and the correlation
 between the
z-component of spin between nearest neighbors, when the auxiliary function
$\varphi(\lambda)$ is/is not singular at $\lambda=1$. For the spin-1 Ising
 model,
we restrict ourselves to the case $h=0$. In
section 5 we present our conclusions.

\section{A Survey of the Calculation of the Grand Potential of a Chain
 Model}

\ \indent\hspace{-3mm} In reference \cite{chain_m} Rojas {\it et al.} 
 obtained, in
the thermodynamic limit,   a closed expression for the grand potential
 of  any translationally invariant chain model with periodic
 boundary condition from an auxiliary function. Here we present a survey of
 the method and
we refer to reference  \cite{chain_m} for further details.

Let us consider a one-dimensional regular lattice (a periodic chain) with
 $N$ sites, so
 that the Hilbert space of the chain model is simply
  ${\mathcal H}^{(N)}=\overset{N}\otimes{\mathcal H}$, ${\mathcal H}$ being
  the irreducible representation at one site, including all its degrees to
 freedom. 
  The dimension of this Hilbert space is 
  ${\rm dim}{\mathcal H}^{(N)}={\rm tr}_N(\lp)$.
  The notation ${\rm tr}_N$ means the trace over all $N$ sites and their
 internal 
  degrees of freedom, e.g. spin.

The grand canonical partition function of a quantum system in the chain
with $N$ sites is given by

\begin{equation}\label{tyl K}
{\mathcal Z}_N(\beta,\mu)={\rm tr}_N({\rm e}^{-\beta\Kp}),
\end{equation}

 \noindent where  $\Kp={\mathbb H}-\mu{\mathbb N}$, with  $\mu$ being the
 chemical potential and ${\mathbb N}$ being an operator that commutes with 
 the Hamiltonian of the system.

Let  ${\bf A}$ be any operator that acts on ${\mathcal H}^{(M)}$ where
 $M\leqslant N$. 
We define $\bra {\bf A}\ket\equiv\frac{{\rm tr}_M({\bf A})}{{\rm
 tr}_M(\lp)}$, 
for any dimension of ${\mathcal H}^{(M)}$.  We call $\bra {\bf A}\ket$ the 
{\it normalized trace} of operator ${\bf A}$. 

Using the definition of normalized trace, eq.\eqref{tyl K} becomes  the
 expansion of
${\cal Z}_N(\beta,\mu)$ around $\beta=0$,
        \begin{equation}\label{tyl K1}
{\cal Z}_N(\beta,\mu)= {\rm tr}_N(\lp)\Big\{1+
\sum_{n=1}^{\infty}(-\beta)^{n}\frac{\bra\Kp^n\ket}{n!}\Big\}.
\end{equation}

In reference \cite{chain_m} we showed that the coefficients
$\bra{\mathbb K}^n\ket$ in eq.\eqref{tyl K} can be written,
for any translationally invariant hamiltonian with interactions between
first neighbors and for arbitrary $n$ as

\begin{equation}\label{TrnK}
\frac{\bra{\Kp}^n\ket}{n!}=\sum_{r=1}^{[n,N]}\sum_{m=r}^{[n,N]}
      \frac{N}{r}\binom{N-m-1}{r-1}
{\rm K}^{(n)}_{r,m}\;.
\end{equation}

 \noindent The notation $[n,N]$ means the ${\rm min}(n,N)$ and
 ${\rm K}_{r,m}^{(n)}$ is defined by 

\begin{equation}\label{Krmn}
{\rm K}_{r,m}^{(n)}\equiv
 {\sump_{\{n_i\}}^{n}} {\sump_{\{m_i\}}^{m}}\prod_{j=1}^{r}{\rm
 K}_{1,m_j}^{(n_j)}\;,
\end{equation}

 \noindent where ${\scriptsize\underset{\{n_i\}}{\overset{n}{\sump}}}$
  means the restriction: $\underset{i=1}{\overset{m}{\sum}}n_i=n$ and
  $n_i\ne 0$ for $i=1,2,..,m$. We also use the notation
 $\{n_i\}\equiv\{n_1,n_2,\dots,n_m\}$ and
 $\{m_i\}\equiv\{m_1,m_2,\dots,m_m\}$.
  The function ${\rm K}_{1,m}^{(n)}$ is defined as

\begin{align}\label{K1mn}
{\rm K}_{1,m}^{(n)}=
{\sump_{\{n_i\}}^{n}}\big\bra\prod_{i=1}^{m}\frac{{\bf
 K}_{i,i+1}^{n_i}}{n_i!}\big\ketg,
\end{align}

\noindent  and each term on the r.h.s. of eq.\eqref{K1mn} corresponds to
 the g-trace of 
an open connected sub-chain.

In the definition of the function ${\rm K}_{1,m}^{(n)}$ we have the g-trace
 which means

{\normalsize
\begin{align}\label{def Kg}
\left\bra\frac{{\bf K}_{{ i_1,i_1+1}}^{n_1}{\bf K}_{i_2,i_2+1}^{n_2}\dots
{\bf K}_{i_m,i_m+1}^{n_m}}{n_1!\;n_2!\;\dots\; n_m!}\right\ketg\equiv 
\frac{1}{n!}\sum_{\mathcal P}\bra {\mathcal P}({\bf K}_{i_1,i_1+1}^{n_1}\,,
         {\bf K}_{i_2,i_2+1}^{n_2}\,,\dots,{\bf K}_{i_m,i_m+1}^{n_m})\ket,
\end{align}}

\noindent where $\sum_{i=1}^{m}n_i=n$ with $n_i\ne 0$ and the indices
$i_k$, $k=1..m$  are distinct among themselves.  The notation 
 $\bra {\mathcal P}({\bf K}_{i_1,i_1+1}\,,{\bf K}_{i_2,i_2+1}\,,\dots,
 {\bf K}_{i_m,i_m+1})\ket$ represents all the distinct permutations of the
  $m$ operators $\{{\bf K}_{i_1,i_1+1}\,, {\bf K}_{i_2,i_2+1}\,,\dots,{\bf
 K}_{i_m,i_m+1}\}$.

We show in reference \cite{chain_m}, that the grand potential
per site ${\cal W}(\beta,\mu)$, in the thermodynamic limit is written as

\begin{align}\label{WTheta1}
{\cal W}(\beta,\mu)&= -\frac{1}{\beta}\left\{\ln({\rm tr}_1({\bf
 1}))+\ln(1+\xi)\right\}.
\end{align}

 \noindent where 

\begin{equation}\label{s(lamd)}
\xi=\sum_{n=0}^{\infty}\frac{{\rm d}^n}{{\rm d}\lambda^n}
\left(\frac{\varphi(\lambda)^{n+1}}{(n+1)!}\right)\bigg|_{\lambda=1}
\end{equation}
\noindent  with $\lambda$ being a parameter, the auxiliary function
 $\varphi(\lambda)$ is
 equal to

\begin{subequations}\label{phigamma}
\begin{align}\label{varphi}
\varphi(\lambda)=\overset{\infty}{\underset{m=1}{\sum}}\frac{\Gamma_{m}}{\lambda^m},
\end{align}

 \noindent and

\begin{align}\label{Gamma_m}
\Gamma_{m}\equiv\overset{\infty}{\underset{n=m}\sum}(-\beta)^n{\rm
 K}_{1,m}^{(n)}.
\end{align}
\end{subequations}

\noindent From eqs.\eqref{WTheta1}-\eqref{Gamma_m} we see that the grand
 potential per 
site, in the thermodynamic limit, can be derived only from the open
 connected sub-chains.
The weight of each sub-chain in the $\beta$-expansion of ${\cal
 W}(\beta,\mu)$ is
 already presented in eq.\eqref{WTheta1}.

\section{The Helmholtz Free Energy of the Spin-1 Ising model}

\ \indent\hspace{-3mm} The hamiltonian of the spin-1 Ising model (with
 single-ion anisotropy) is:

\begin{equation}\label{HIsing}
{\mathbb H} = \sum_{j=1}^{N}
\Delta S^{z}_{j}S^{z}_{j+1}-h S^{z}_{j} + D (S^{z}_{j})^2 ,
\end{equation}

\noindent where $h$ is the external magnetic field in the $z$-direction,
 while
the parameters $\Delta$ and $D$ are  the exchange and single-ion
anisotropy, respectively. The chain has $N$ sites, and
periodic boundary conditions are assumed.

It is worth noticing that the partition function for the model above may
be obtained from the transfer matrix approach by calculating the
eigenvalues of the hamiltonian (\ref{HIsing}). These
eigenvalues are solutions of a polynomial of third degree.
Nevertheless, for arbitrary values of the parameters
$\Delta$, $h$ and $D$, such an approach only gives numerical
solutions for the thermodynamic functions, which
allow us to verify the applicability of the method presented in reference
\cite{chain_m} to one exactly solvable limiting case of a
non-integrable model\cite{takahashi_L}. The  validity of the present
approach has already  been verified for the integrable XXZ Heisenberg
spin-1/2 model\cite{chain_m} and its limiting cases\cite{BJP}.

In order to obtain the Helmholtz free energy, we must  calculate
at first its functions ${\rm H}_{1,m}^{(n)}$. Due to the
fact that all terms that contribute to the hamiltonian of the Ising model
commute among themselves, we may write

\begin{align} {\rm
 H}_{1,m}^{(n)}=\sump_{\{n_i\}}^{n}\Big\bra\prod_{i=1}^{m}
\frac{H_{i, i+1}^{n_i}}{n_i!}
\Big\ket, \hspace{1cm} {\rm with } \hspace{0.3cm}  m\le n.
\end{align}

\noindent The sums in  $\Gamma_m$ (see
eq.(\ref{Gamma_m})),   including the restricted sums over the indices
$\{m_i\}$, are more tractable if we recognize that
they can be substituted,
in the thermodynamic limit, by $m$ independent sums
with each index $\{m_i \}$ varying from $1$ to $\infty$\cite{chain_m}.  For
 this
model, $\Gamma_m$ becomes

\begin{equation}
\Gamma_m = \sum_{n_1,n_2, \cdots,  n_{m} =1 }^\infty \Big\bra
\prod_{i=1}^{m}\frac{(-\beta)^{n_i} H_{i, i+1}^{n_i}}{n_i !} \Big\ket\,
\end{equation}

 \noindent where $ H_{i, i+1} = \Delta S_i^z S_{i+1}^z + B_i^z $,
with $B_i^z \equiv -h S_i^z + D(S_i^z)^2$. The two operators in $ H_{i,
 i+1}$
commute; consequently, we can apply Newton's
multinomial formula to obtain  the expansion of $ H_{i, i+1}^{n_i}$,

\begin{equation}   \label{Hni}
H_{i, i+1}^{n_i} = \sum_{j_i=0}^{n_i}
 \begin{pmatrix} n_i \cr j_i\cr \end{pmatrix}
(\Delta)^{n_i - j_i} (S_i^z)^{n_i - j_i} (B_i^z)^{j_i} (S_{i+1}^z)^{n_i -
 j_i},
\end{equation}

\noindent where $\begin{pmatrix} n_i \cr j_i\cr \end{pmatrix}$ are the
 multinomial
coefficients. After averaging over the space sites from $i=1$ up to
$i= m+1$, we obtain

\begin{eqnarray} \label{GamaSomado}
\Gamma_m &  =&  \frac{1}{3} \prod_{i=1}^{m} \left(
     \frac{1}{3}  \sum_{n_i =1}^{\infty} \frac{(-\beta)^{n_i}}{n_i!}
 \sum_{j_i =0}^{n_i}   \begin{pmatrix} n_i \cr j_i\cr \end{pmatrix}
\Delta^{n_i - j_i} \left( (D-h)^{j_i} + (-1)^{n_i-j_i} (D+h)^{j_i}  
 \right)
     \right)  \times  \nonumber \\
&& \nonumber \\
%
 && \hspace{3cm} \times
[1 + (-1)^{n_m - j_m} + \delta_{j_m, n_m}].
\end{eqnarray}

It is tedious but simple to show that

\begin{eqnarray}  \label{Sn}
 \frac{1}{3}  \sum_{n_m =1}^{\infty} \frac{(-\beta)^{n_m}}{n_m!}
   && \hspace{-0.7cm}   \sum_{j_m =0}^{n_m}   \begin{pmatrix} n_m \cr
 j_m\cr \end{pmatrix}
      \Delta^{n_m - j_m} \left( (D-h)^{j_m} + (-1)^{n_m-j_m} (D+h)^{j_m}
 \right)
      \times  \nonumber \\
   & & \hspace{-2cm}  \times
[1 + (-1)^{n_m - j_m} + \delta_{j_m, n_m}] = (-1)^{n_{m-1} - j_{m-1}} a_1 +
 b_1,
\end{eqnarray}

\noindent where the constants $a_1$ and $b_1$ have been defined as

\begin{equation}  \label{a1b1}
 a_1 \equiv  p^{+} + q^{-} + r^{-}
        \hspace{1cm} {\rm and} \hspace{1cm}
 b_1 \equiv  p^{-} + q^{+} + r^{+}
\end{equation}

\noindent and

\begin{align}\label{pqr}
p^{\pm} \equiv \frac{1}{3}\Big({\rm e}^{\pm\beta(\Delta-(h \pm D))}-1\Big),
\quad
 q^{\pm} \equiv  \frac{1}{3}\Big({\rm e}^{\pm\beta(\Delta+ (h \mp
D)}-1\Big)\quad\text{and}\quad
 r^{\pm} \equiv  \frac{1}{3}\Big({\rm e}^{\pm\beta( h \mp D)}-1\Big)\,.
 \end{align}

Performing the sums in the  product on the r.h.s. of eq.(\ref{GamaSomado}),
 we get a recursive solution

\begin{equation}
\Gamma_m = \frac{ a_m + b_m}{3},
\end{equation}

\noindent where

\begin{equation}\label{21}
a_i  = b_{i-1}\, p^{+} + a_{i-1}\,  q^{-}
\hspace{0.5cm} {\rm and} \hspace{0.5cm}
b_i  =   b_{i-1}\,  p^{-} + a_{i-1} \, q^{+},
\hspace{0.5cm}  i=2, 3, \cdots, m.
\end{equation}

 Eq.(\ref{varphi}) gives us the relation between the auxiliary function
$\varphi(\lambda)$ and $\Gamma_m$ which, for this particular model, is

\begin{equation}
\varphi(\lambda) = \frac{1}{3} \sum_{m=1}^{\infty} \frac{a_m}{\lambda^m} +
      \frac{1}{3} \sum_{m=1}^{\infty} \frac{b_m}{\lambda^m}.
\end{equation}

\noindent Defining $\varphi(\lambda) \equiv \phi_a(\lambda) +
 \phi_b(\lambda)$ with

\begin{equation} \label{23}
 \phi_a(\lambda) \equiv  \frac{1}{3} \sum_{m=1}^{\infty}
 \frac{a_m}{\lambda^m}
\hspace{1.3cm} {\rm and}  \hspace{1.3cm}
 \phi_b(\lambda) \equiv  \frac{1}{3} \sum_{m=1}^{\infty}
 \frac{b_m}{\lambda^m},
\end{equation}

\noindent we may use the relations (\ref{21})  to rewrite $\phi_a(\lambda)$
and $\phi_b(\lambda)$ as

\begin{subequations}   \label{24}

\begin{equation}
\phi_a(\lambda) = \frac{a_1}{3 \lambda} + \frac{p^{+}}{\lambda}
 \phi_b(\lambda)
                + \frac{q^{-}}{\lambda} \phi_a(\lambda)
\end{equation}

\noindent and

\begin{equation}
\phi_b(\lambda) = \frac{b_1}{3 \lambda} + \frac{p^{-}}{\lambda}
 \phi_b(\lambda)
                + \frac{q^{+}}{\lambda} \phi_a(\lambda)
\end{equation}

\end{subequations}

\noindent which yields

\begin{align}\label{phifinal}
\varphi(\lambda)=\frac{1}{3}\frac{\big(p^{+}+p^{-}+q^{+}+q^{-}+r^{+}+r^{-}\big)
\lambda+2\big(p^{+}q^{+}-p^{-}q^{-}\big)+r^{+}\big(p^{+}-q^{-}\big)+r^{-}
\big(q^{+}-p^{-}\big)}{\lambda^2-\big(p^{-}+q^{-}\big)\lambda-\big(p^{+}
q^{+}-p^{-}q^{-}\big)}.
\end{align}

\noindent The Helmholtz free energy of the Ising model is then obtained by
 substituting
eq.\eqref{phifinal}  in eq.(\ref{WTheta1}), that is,

\begin{align}\label{Ising_limit}
{\cal W}(\beta)&= - \frac{1}{\beta}\Big\{\ln(3)+\ln\Big(1+
\sum_{n=0}^{\infty}\frac{{\rm d}^n}
{{\rm d}\lambda^n}\left(\frac{\varphi(\lambda)^{n+1}}{(n+1)!}
\right)\bigg|_{\lambda=1}\Big)\Big\},
\end{align}

\noindent and from it we derive a  $\beta$-expansion of  ${\cal W}(\beta)$
 with
arbitrary  value of $n$  through an algebraic computational language
such as {\tt MAPLE}.

Depending on the values of $\Delta$, $h$ and $D$, the function $\varphi
 (\lambda)$
may be singular at $\lambda=1$ for some values of $\beta$. This is also
 true for
powers of this function at $\lambda=1$ which contribute to
 eq.(\ref{Ising_limit}).
However, we know that there is no phase transition at finite $\beta$ in any
 one-dimensional Ising model. Consequently, those singularities in
$\varphi (1)$ (i.e., $\varphi (\lambda=1)$) must be non-physical. We point
out that a similar behavior is shown by the auxiliary function for the
spin-1/2 Ising model (see reference \cite{BJP}).  A question yet to be
answered is whether the presence of non-physical singularities in $\varphi(1)$
could indicate a poorer convergence of
 high temperature expansions of the thermodynamical quantities. In the
next section we compare the high temperature expansions for the Ising
models with spin-1/2 and spin-1, when their respective auxiliar functions
have a singularity at $\lambda=1$, and when they do not.

\section{Comparison of the  High  Temperature Expansions  of \\
 $s=\frac{1}{2}$  and $ s=1$ Ising models}

\ \indent\hspace{-3mm} The auxiliary function $\varphi (\lambda)$ appears
 for
the first time in the
literature in our reference \cite{chain_m}. Therefore
calculating it for
exactly solvable models is useful to the understanding of its properties,
including the impact of its singularities on the convergence
of the $\beta$-expansion of thermodynamic quantitities.

The function  $\varphi_1 (\lambda)$  of the spin-1 Ising   model
(see eq.(\ref{phifinal})) can be written  as

\begin{subequations}  \label{30}

\begin{equation}  \label{30a}
\varphi_1 (\lambda) = \frac{A(\beta)}{\lambda - \lambda_{+}}
                   + \frac{B(\beta)}{\lambda - \lambda_{-}},
\end{equation}

\noindent where  $\lambda_{\pm}$ are the roots of the polynomial of second
degree in $\lambda$ in the denominator of eq.(\ref{phifinal}), namely,

\begin{equation}   \label{30b}
\lambda_{\pm} = \frac{1}{2} \left[q^{-} + p^{-} \pm
 \sqrt{(q^{-} - p^{-})^2 + 4 q^{+} p^{+}} \right] .
\end{equation}

\end{subequations}

\noindent The constants $A(\beta)$ and $B(\beta)$ are easily obtained
 substituting
eq.(\ref{30a}) in eq.(\ref{phifinal}).

For the case $h=0$, eqs.(\ref{30}) reduce to

\begin{subequations} \label{31}

\begin{equation}  \label{31a}
\varphi_1^{(0)} (\lambda) = \frac{A(\beta)} { \lambda -(p^{+} + p^{-})},
\end{equation}

\noindent with

\begin{equation} \label{31b}
A(\beta) = \frac{2}{3} (p^{+} + p^{-} + r^{-}).
\end{equation}

\end{subequations}

For $h=0$, the type of $\lambda$-dependence in
eq.(\ref{31a}) gives rise to an alternating series to the Helmholtz free
energy, in the variable $A(\beta)/(1-\lambda_{+})^2$.

 For $ -D< \Delta <D$ and  $D>0$, the function
$\varphi_1^{(0)} (1)$ has no singularity; otherwise, there is one
real and positive value of $\beta$
for which it is singular.

The auxiliary function (\ref{31a}) has the same $\lambda$-dependence as
the auxiliary function of the spin-1/2 Ising model, that was summed up
in reference \cite{BJP}.
Substituting eq.(\ref{31a}) in eq.(\ref{phifinal}), and following
the steps described in \cite{BJP}, we obtain
the Helmholtz free energy of the spin-1 Ising model  at $h=0$,

\begin{equation} \label{32}
 {\cal W}_1 (\beta) = - \frac{1}{\beta}
\ln \left[ \frac{3}{2} \left(1 + p^{+} + p^{-}
+ \sqrt{(1 -  p^{+} - p^{-})^2 +\frac{3}{8} ( p^{+} + p^{-} + r^{-}) }
 \right) \right],
\end{equation}

\noindent valid for finite values of $\beta$.
This result coincides with the one derived from the transfer matrix
approach, as well as with those obtained by numerical analysis.

\vspace{0.2cm}

We derived in reference \cite{BJP} the auxiliary function
 $\varphi_{\frac{1}{2}} (\lambda)$ for
the spin-1/2 Ising model. For arbitrary value of the external magnetic
field $h$ we got

\begin{equation}  \label{33}
\varphi_{\frac{1}{2}} (\lambda) = \frac{e^{2\beta h} + e^{2\beta
 (h+\Delta)} -2} {4}
\left(  \frac{1}{\lambda - \frac{e^{\beta h} -1}{2}  }  \right).
\end{equation}

\noindent Only for $h\not=0$ his function has one real and positive value
of $\beta$ where $\varphi_{\frac{1}{2}} (1)$ is singular.
The Helmholtz free energy of the spin-1/2 Ising model, for arbitrary
value of $h$ is\cite{corrBJP}

\begin{equation}   \label{34}
{\cal W}_{\frac{1}{2}} (\beta) = - \frac{1}{\beta}
\ln\left[ e^{-\frac{\beta \Delta}{4}} \cosh(\beta h)
+ \sqrt{  e^{-\frac{\beta\Delta}{2}} \sinh^2(\beta h) +
      e^{\frac{\beta\Delta}{2} }} \;  \right],
\end{equation}

\noindent that is also valid for finite values of $\beta$.

In both equations (\ref{34}) and (\ref{32}), the respective Helmholtz
free energies are $\beta$-expansions of infinite range.
We are interested in the impact of the singularities of
$\varphi(\lambda)$ on the rapidity of convergence of the $\beta$- expansion
 of thermodynamical quantities. For both spin=1 and spin-1/2
models,   we have taken the specific heat $C_v$ and the z-component of spin
correlation $<S^z_i S^z_{i+1}>$ between nearest neighbors as examples, and
have chosen two suitables sets of values for the parameters
$(\Delta,D,h)$ so that the singular and non-singular $\varphi$ cases can be
compared, as far as the rapidity of convergence of $\beta$-expansions is
concerned. In what follows, {\em approximate curves} will refer to
those obtained by truncation of the $\beta$-expansion at 80th order in
$\beta$, in contrast to the {\em exact curves}.

For the spin-1 Ising model, we
have taken the first set of conditions to be $(\Delta=1,D=1,h=0)$, which
yields a non-singular $\varphi_1^{(0)}(1)$;  by promoting a slight
variation in $D$, a second set
$(\Delta=1,D=0.99,h=0)$ can be defined, yielding a singular
$\varphi_1^{(0)}(1)$. For both sets, exact and approximate curves
of the specific heat $C_v$ are shown in figure 1 and 2.
We expect that both
exact curves be slightly apart from each other, in some finite interval of
sufficiently small $\beta>0$. They differ, indeed, by a relative error
inferior to $1\%$, for $0\leq\beta<2.5$. If we compare the behavior of the
corresponding approximate curves, however, we observe
that convergence in the singular $\varphi$ case is much worse than that of
the non-singular case. In the non-singular case, the approximate and exact
curves differ by less than $1\%$ for $0\leq\beta<1.16$, whereas in the
singular case they differ by less than $1\%$ for a much smaller interval
$0\leq\beta<0.54$.
Similar behavior can be observed in the
correlation function $<S^z_i S^z_{i+1}>$ (see figures 3 and 4).
Exact curves in both singular and non-singular cases differ by less than
$2\%$ in the interval $0\leq\beta<1.5$; however, in the non-singular case
the approximate and exact curves differ by less than $1\%$ in
$0\leq\beta <1.23$, whereas in the singular case this interval is much
shorter, namely, $0\leq\beta <0.58$.

Similar discussion can be carried out for the spin-1/2 Ising model. Here, the conditions
$(\Delta=-1,D=1,h=0)$ and $(\Delta=-1,D=1,h=0.01)$ yield non-singular
and singular cases, respectively. (Observe the slight variation in $h$,
only.) For both sets, exact and approximate curves
of the specific heat $C_v$ are shown in figure 5 and 6.
The exact curves differ by less than
$1\%$ in the interval $0\leq\beta<2.6$.
 In the non-singular case, the approximate and exact
curves differ by less than $1\%$ for $0\leq\beta<5.2$, whereas in the
singular case the same difference holds for a much smaller
interval $0\leq\beta<1.8$.
The same behavior
is also exhibited by the correlation function $<S^z_i S^z_{i+1}>$
(see figures 7 and
8). Exact curves in both singular and non-singular cases differ by less
than $0.05\%$ in the interval $0\leq\beta<1.8$; however, in the non-singular
case the approximate and exact curves differ by less than $1\%$ in
$0\leq\beta<1.23$, whereas in the singular case this interval is much
shorter, $0\leq \beta<0.58$.

Those results indicate that the presence of singularities in
the auxiliary function $\varphi$ yields poor convergence of
$\beta$-expansions of thermodynamical functions.



\psfrag{0.1}{\small 0.1}
\psfrag{0.2}{\small 0.2}
\psfrag{0.3}{\small 0.3}
\psfrag{0.4}{\small 0.4}
\psfrag{0.5}{\small 0.5}
\psfrag{0.6}{\small 0.6}
\psfrag{0.8}{\small 0.8}
\psfrag{1}{\small 1}
\psfrag{0.12}{\small 0.12}
\psfrag{0.08}{\small 0.08}
\pf{0.18} \pf{0.16} \pf{0.14}
\pf{0.05} \pf{0.04} \pf{0.02}
\pf{0.5} \pf{0.06} \pf{-0.2}
\pf{-0.4} \pf{-0.2} \pf{1.2}
\pf{0.5}\pf{-0.5}
\pf{-0.1}

\psfrag{Titulo}{\small Specific heat $(s=1)$}
\psfrag{h}{$h=0$}
\psfrag{EixoX}{$\beta$}
\psfrag{EixoY}{\ \ \ $C_v$}
\begin{figure}[htb]
	\begin{minipage}{8cm}
	\includegraphics{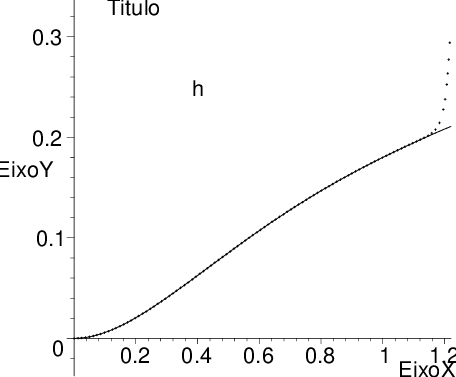}
	\caption{The full line is the exact result of the specific heat
		for $\Delta=1$ and $D=1$. The
		dotted line is its expansion up to order n=80 in
		$\beta \in [0, 1.22].$}
	\end{minipage}

\psfrag{Titulo}{\hspace{3mm}\small Specific heat $(s=1)$}
\psfrag{h}{$h=0$\hspace{-4cm}$C_v$}
\psfrag{EixoX}{\hspace{-3mm}$\beta$}
\psfrag{EixoY}{}
\psfrag{\260.1}{\small -0.1}
\psfrag{0.1}{\small 0.10}
\vspace{1.5cm}
	\begin{minipage}{8cm}
	\includegraphics{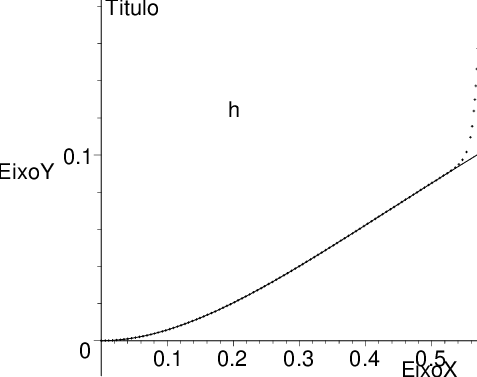}
	\caption{The full line is the exact result of the
		specific heat for $\Delta=1$ and $D=0.99$. The
		dotted line is its expansion up to order n=80 in
		$\beta \in [0, 0.57].$}
	\end{minipage}

\psfrag{Titulo}{\small
	\begin{minipage}{6cm}
	Correlation between nearest \\ neighbors $(s=1)$
	\end{minipage}
	}
\psfrag{h}{$h=0$}
\psfrag{EixoX}{\hspace{3mm}$\beta$}
\psfrag{EixoY}{
	\hspace{3mm}
	\rotatebox{90}{$<S^z_i S^z_{i+1}>$}
	}
 \psfrag{\260.5}{\small -0.5}
\vspace*{-19.4cm}\hspace{8.5cm}
	\begin{minipage}{8cm}
	\includegraphics{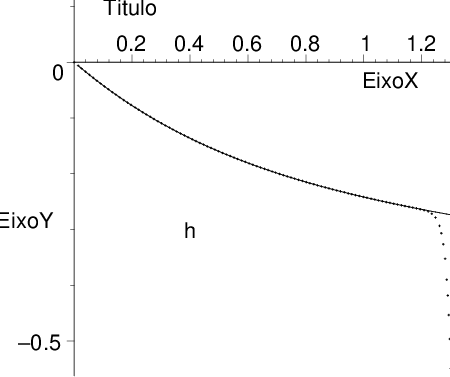}
	\caption{The full line is the exact result of
		the correlation between nearest neighbors for
		$\Delta=1$ and $D=1$. The
		dotted line is its expansion up to order n=80 in
		$\beta \in [0, 1.3].$ }
	\end{minipage}

\psfrag{Titulo}{\small
	\begin{minipage}{6cm}
	Correlation between nearest \\ neighbors $(s=1)$
	\end{minipage}
	}
\psfrag{h}{$h=0$}
\psfrag{EixoX}{$\beta$}
\psfrag{EixoY}{
	\hspace{1cm}
	\rotatebox{90}{$<S^z_i S^z_{i+1}>$}
	}
 \psfrag{\260.1}{\small -0.1}
\vspace{1.5cm}\hspace{8.5cm}
	\begin{minipage}{8cm}
	\includegraphics{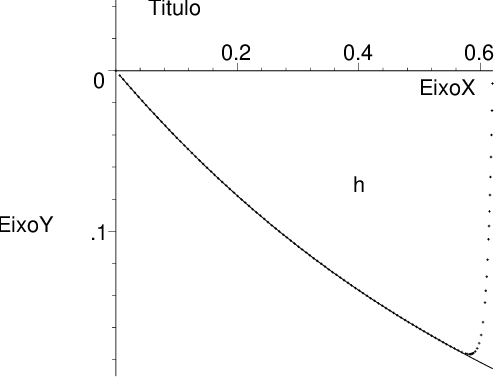}
	\caption{The full line is the exact result of
		the correlation between nearest neighbors for
		$\Delta=1$ and $D=0.99$. The
		dotted line is its expansion up to order n=80 in
		$\beta \in [0, 0.62].$ }
	\end{minipage}
\end{figure}


\psfrag{Titulo}{\hspace{3mm}\small Specific heat $(s=1/2)$}
\psfrag{h}{$h=0$\hspace{-3.3cm}$C_v$}
\psfrag{EixoX}{$\beta$}
\psfrag{EixoY}{}

\begin{figure}[htb]
\pf{2} \pf{4}
\psfrag{0.4}{}
	\begin{minipage}{8cm}
	\includegraphics{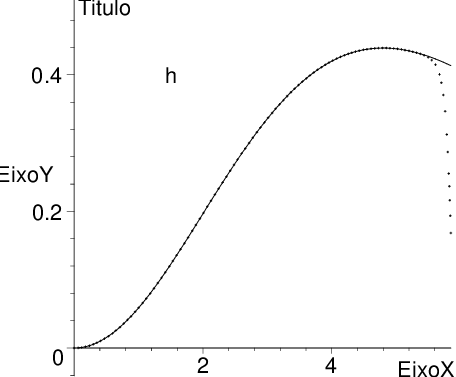}
	\caption{The full line is the exact result of the specific heat
		for $\Delta=-1$ and $D=1$. The
		dotted line is its expansion up to order n=80 in
		$\beta \in [0, 5.85].$}
	\end{minipage}

\psfrag{Titulo}{\hspace{4mm}\small Specific heat $(s=1/2)$}
\psfrag{h}{\hspace{5mm}$h=0.01$\hspace{-3.3cm}$C_v$}
\psfrag{EixoX}{$\beta$}
\psfrag{EixoY}{}
\psfrag{0.2}{}
\psfrag{0.4}{}
\psfrag{0.8}{}
\psfrag{1}{}
\psfrag{1.4}{}
\psfrag{1.6}{}
\pf{1.8}

\vspace{1.5cm}
	\begin{minipage}{8cm}
	\includegraphics{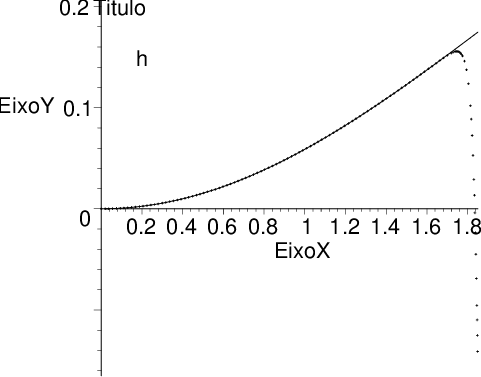}
	\caption{The full line is the exact result of the
		specific heat for $\Delta=-1$ and $D=1$, but for
		$h=0.01$. The dotted line is its expansion
		up to order n=80 in
		$\beta \in [0, 1.85].$}
	\end{minipage}

\psfrag{Titulo}{\small \hspace{3mm}
	\begin{minipage}{6cm}
	Correlation between nearest \\ neighbors $(s=1/2)$
	\end{minipage}
	}
\psfrag{h}{$h=0$}
\psfrag{EixoX}{$\beta$}
\psfrag{EixoY}{
	\hspace{3.5mm}
	\rotatebox{90}{$<S^z_i S^z_{i+1}>$}
	}
 \pf{6}
\vspace*{-19.4cm}\hspace{8.5cm}
	\begin{minipage}{8cm}
	\includegraphics{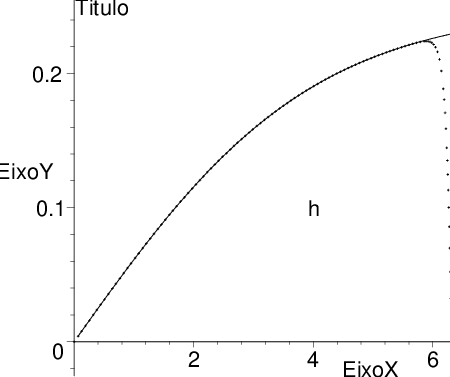}
	\caption{The full line is the exact result of
		the correlation between nearest neighbors for
		$\Delta=-1$ and $D=1$. The
		dotted line is its expansion up to order n=80 in
		$\beta \in [0, 6.3].$}
	\end{minipage}

\psfrag{Titulo}{\small \hspace{4mm}
	\begin{minipage}{6cm}
	Correlation between nearest \\ neighbors $(s=1/2)$
	\end{minipage}
	}
\psfrag{h}{$h=0.01$}
\psfrag{EixoX}{$\beta$}
\psfrag{EixoY}{
	\hspace{3mm}
	\rotatebox{90}{$<S^z_i S^z_{i+1}>$}
	}
 \psfrag{\260.1}{\small -0.1}
 \psfrag{0.1}{}
\vspace{1.5cm}\hspace{8.5cm}
	\begin{minipage}{8cm}
	\includegraphics{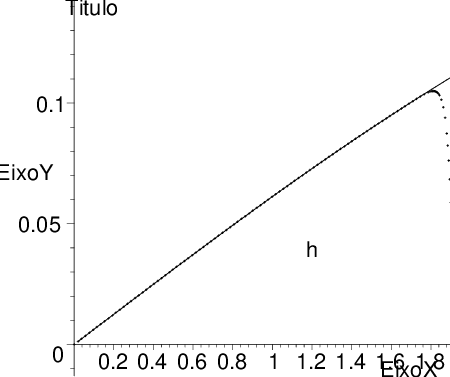}
	\caption{The full line is the exact result of
		the correlation between nearest neighbors for
		$\Delta=-1$ and $D=1$ but for $h=0.01$. The
		dotted line is its expansion up to order n=80 in
		$\beta \in [0, 1.9].$ }
	\end{minipage}
\end{figure}

\section{Conclusions}

\ \indent\hspace{-3mm} In reference \cite{chain_m} we obtained,
 in the thermodynamical limit, a
closed analytical series for the grand potential for any unidimensional
chain model with periodic boundary conditions, from the high-temperature
expansion of the cumulant method \cite{domb}. In order to derive the
grand potential, we must calculate the auxiliary function
$\varphi(\lambda)$ for the particular model of interest. Recently, the
method presented by Rojas {\em et al.} was applied successfully to the
spin-1/2 XXZ Heinsenberg model \cite{chain_m} and its limiting
 cases\cite{BJP}.
In the present work we consider the exactly solvable spin-1 Ising model,
whose numerical solutions can be easily attained and used in the
verification of the correctness of the results derived from reference \cite{chain_m},
as well
as in the study of the properties of the auxiliary function
$\varphi(\lambda)$ of this model. From eqs. (\ref{phifinal}) and
(\ref{Ising_limit}) we obtain the $\beta$-expansion for the Helmholtz
free energy ${\cal W}(\beta)$ for arbitrary values of the parameters
$\Delta$, $D$ and $h$.
The analyticity of this $\beta$-expansion allows its use as input
to the perturbative study of thermodynamical properties
of uniaxial Ising-like models.
For one set of values of the parameters, and $n=40$
(where $n$ is the leading order of the $\beta$-expansion) we plotted
those expansions and confirmed that our results match the numerical
results for this thermodynamic function, in an interval of $\beta$,
including situations in which $h\neq 0$. For $h=0$ we are able to sum
the terms in the expansion (\ref{Ising_limit}) of ${\cal W}_1(\beta)$
for arbitrary values of $\Delta$ and $D$ at any finite $\beta$. In this
case, the auxiliary function $\varphi_1^{(0)}(\lambda)$ is given by eq.
(\ref{31}), and it has no singularity at $\lambda=1$ for $-D<\Delta<D$
and $D>0$; otherwise, there is a value of $\beta$ for which
$\varphi_1^{(0)}(\lambda)$ is singular. Such singularity for the spin-1
model is of the same type as that of spin-1/2 Ising model \cite{BJP}. In
both models the singularities are non-physical, since no unidimensional
Ising model has phase transitions at finite $\beta$. In order to see if the
existence of a singularity in the auxiliary functions of both models
could influence the rapidity of convergence of the $\beta$-expansion of
the thermodynamic functions, for each model we considered two distinct
sets of parameter values: in one of them the respective auxiliary
function is singular, and in the other one it has no singularities, at
$\lambda=1$. For each model, the two sets differ by the value of only
one parameter, and the difference in value is very small, in such a way
that the results for those sets could be related to one another
by perturbation theory,
at least in a finite range of $\beta$.

In the $\beta$-expansion of thermodynamic functions we kept $n=80$ for
both models and both cases (singular and non-singular $\varphi^\prime$s
, spin-1/2 and spin-1). These
results strongly indicate that the presence of singularities in the
auxiliary function allows one to infer that the $\beta$-expansions will
have poor convergence, even though the series (\ref{phifinal}) has
infinite range in $\beta$.

Our analytical results at $h=0$ can be used as the starting point for
analytical perturbative expansions for $h \neq 0$, and also for other
models.

\vskip .5cm

\centerline{\bf \Large Acknowledgments}\vskip .3cm

W. A. M-M. acknowledges FAPEMIG and CNPq for financial support. O. R.
thanks CLAF for financial support. E.V.C.S. thanks CNPq for financial
support. M.T.T and S.M. de S. thank CNPq for partial financial support.
S.M. de S. thanks FAPEMIG for partial financial support. M.T.T. thanks
FAPERJ for partial financial support.



\begin{thebibliography}{XX}
\bibitem{corrBJP} There is a misprint in reference \cite{BJP}: in the
 expression of the Helmholtz free energy, the constant $\Delta$ should be
 replaced by $\frac{\Delta}{2}$.
\end{thebibliography}
\end{document}